\documentclass[a4paper,11pt]{article}
\usepackage{pos}

\title{Comprehensive power spectral density analysis of the \textit{Fermi}-LAT $\gamma$-ray light curves of selected blazars}
\ShortTitle{PSD analysis of \textit{Fermi}-LAT $\gamma$-ray LCs of blazars}

\author*[a]{Natalia \.{Z}ywucka}
\author[b]{Mariusz Tarnopolski}
\author[b]{Volodymyr Marchenko}
\author[c]{Javier Pascual-Granado}

\affiliation[a]{Centre of Space Research, North-West University, Potchefstroom, South Africa}

\affiliation[b]{Astronomical Observatory, Jagiellonian University, Krak\'ow, Poland}

\affiliation[c]{Instituto de Astrof\'isica de Andaluc\'ia -- CSIC, Granada, Spain}

\emailAdd{n.zywucka@oa.uj.edu.pl}
\emailAdd{mariusz.tarnopolski@uj.edu.pl}

\abstract{We present the results of the \textit{Fermi}-Large Area Telescope (LAT) light curve (LC) modelling of selected blazars: six flat spectrum radio quasars (FSRQs) and five BL Lacertae (BL Lacs). All objects have densely sampled and long-term LCs, over 10 years. For each blazar we generated three LCs with 7, 10, and 14 days binning, using the latest LAT 8-year source catalog and binned analysis provided within the \textsc{fermipy} package. The LCs were modelled with several tools: the Fourier transform, the Lomb-Scargle periodogram (LSP), the autoregressive moving average (ARMA), the fractional autoregressive integrated moving average, the continuous-time autoregressive moving average (CARMA) processes, the Hurst exponents ($H$), the $\mathcal{A}-\mathcal{T}$ plane, and the wavelet scalogram. The power law indices $\beta$ calculated from the Fourier and LSP modelling are consistent with each other. Many objects yield $\beta\simeq 1$, with PKS 2155--304 even flatter, but some are significantly steeper, e.g. Mrk 501 and B2 1520+31. The power law PSD is indicative of a self-affine stochastic process characterised by $H$, underlying the observed variability. Several algorithms for the $H$ estimations are employed. For some objects we found $H>0.5$, indicating long-term memory. We confirm a quasi-periodic oscillation (QPO) in the PKS~2155$-$304 data with the period of $612\pm 42$~days at a $3\sigma$ significance level, but do not detect any QPOs in other objects. The ARMA results give in general higher orders for 7 days binned LCs and lower orders for 10 and 14 days binned LCs, implying temporal variations in the LCs are consistently captured by the fitted models. CARMA modelling leads to featureless PSDs. The recently introduced $\mathcal{A}-\mathcal{T}$ plane allows us to successfully classify the PSDs based on the LCs alone and clearly separates the FSRQ and BL Lac types of blazars.}

\FullConference{%
  *** High Energy Astrophysics in Southern Africa (HEASA2021) ***\\
  *** 13 - 17 September, 2021 ***\\
  *** Online ***
}

\begin{document}
\maketitle

\section{Introduction}

Blazars form a peculiar class of active galactic nuclei (AGNs). They are radio-loud objects pointing their relativistic jets towards an observer and  having a non-thermal continuum along the entire electromagnetic spectrum. One of their properties is rapid variability in different energy bands, lasting from months down to minutes. Generally, blazars are split into two groups, namely flat spectrum radio quasars (FSRQs) and BL Lacertae (BL Lac) objects, based on characteristics visible in their optical spectra, wherein FSRQs have prominent emission lines and BL Lacs are featureless or with weak lines only.

We aim to conduct a comprehensive analysis of the temporal properties of the light curves (LCs) of blazars in our sample, in particular constraining the shape and features of the power spectral density (PSD) to look for short- and long-lasting features. This can be used to establish the variability regions and physical processes responsible for variability and, in some cases, to estimate the black hole (BH) mass of blazars. First of all, we employ standard and well established methods to study characteristics of PSDs, such as breaks, which can point to regions responsible for variability. Subsequently, we want to verify the existence of quasi-periodic oscillations (QPO) which is defined as \textit{''concentration of variability power over a limited frequency range''} \citep{Vaughan2005}. This can shed additional light on the structure of blazars. 

\section{The sample}

\begin{figure}[!h]
\centering
\includegraphics[width=0.8\textwidth]{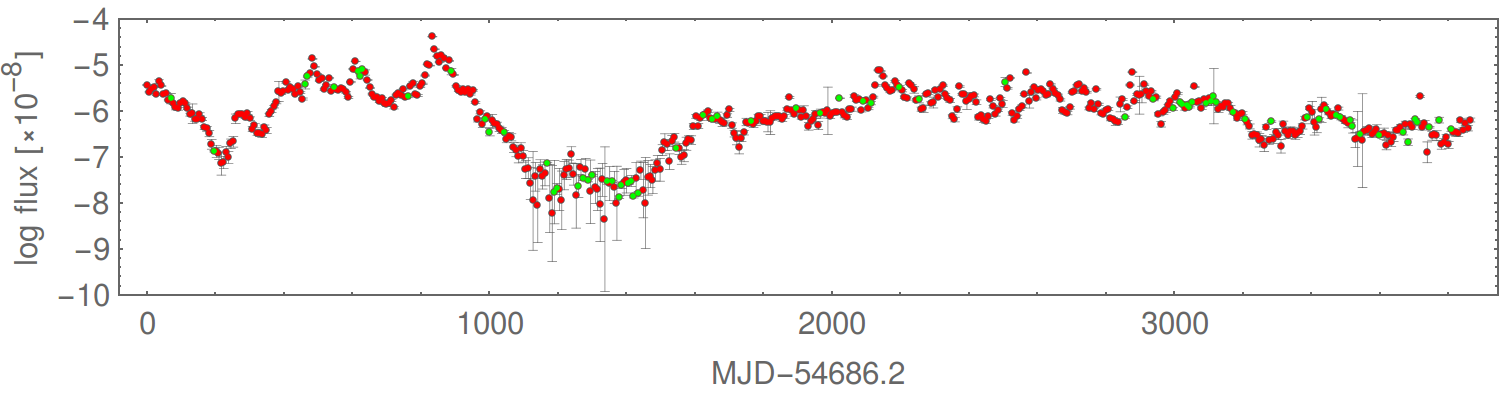}
\includegraphics[width=0.8\textwidth]{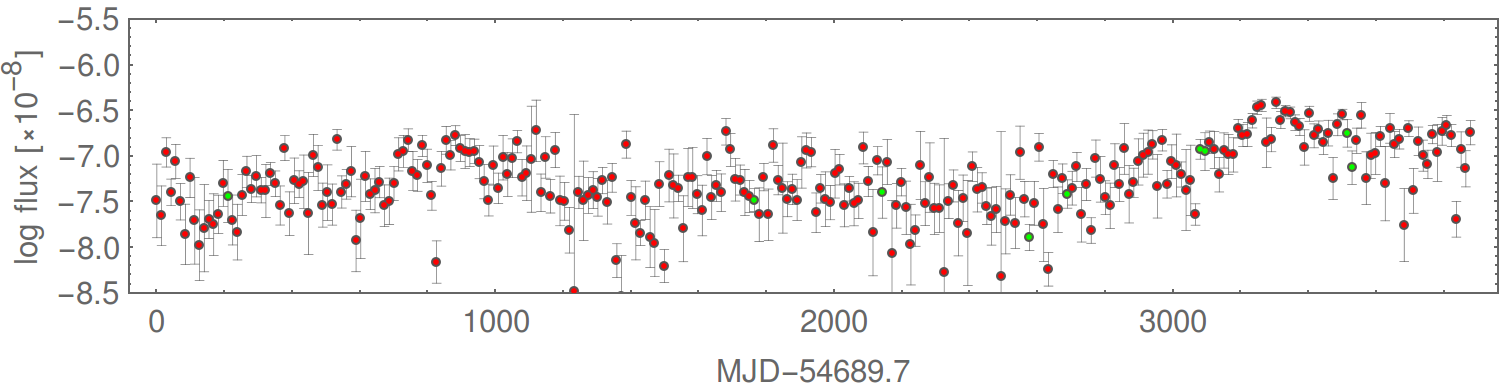}
\caption{Logarithmic LCs of 3C~454.3 (top panel) and TXS 0506+056 (bottom panel). The red points are the observed data and the green points are the interpolations done with MIARMA. }
\label{LCs}
\end{figure}

We analysed with a number of techniques the \textit{Fermi}-LAT $\gamma$-ray LCs of 11 well known blazars, including six FSRQs, PKS 1510$-$089, 3C~279, B2~1520+31, B2~1633+38, 3C~454.3, and PKS 1830$-$211, and five BL Lacs, Mrk~421, Mrk~501, PKS~0716+714, PKS 2155$-$304, and TXS 0506+056. We performed a standard binned maximum likelihood analysis\footnote{\url{https://fermi.gsfc.nasa.gov/ssc/data/analysis/scitools/binned\_likelihood\_tutorial.html}}, using the \textsc{Fermitools}\footnote{\url{https://github.com/fermi-lat/Fermitools-conda/wiki}} and the \textsc{fermipy}\footnote{\url{https://fermipy.readthedocs.io/en/latest/}} packages. In this analysis, we used data from the \textit{LAT 8-year Source Catalog} \citep[4FGL; ][]{Fermi2019}, spanning the time range of 239557417---577328234 MET, which corresponds to $\sim$11 years, in the energy range of 100~MeV up to 300~GeV. We generated a set of three LCs for each blazar, using 7, 10, and 14~days binning. Only the observations with the test statistic $TS>25$ (significance of $\gtrsim 5\sigma$) were taken into account. Eventually, 33 LCs were generated and then analysed. Since the fraction of missing points can reach 13\%, we utilised the method of interpolation by autoregressive and moving average \citep[MIARMA; ][]{granado2015}. Figure~\ref{LCs} presents examples of the LCs. 

\section{Methodology}

This proceedings is based on the \cite{Tarnopolski2020} publication and the methodology is described there in details. \\

\textbf{Fourier transform and Lomb-Scargle periodogram} \citep{Lomb1976,Scar82} are methods to generate a global PSD for a given time series, to calculate indices, to check their relation to the coloured noise, and to investigate global components of the LCs, such as short- and long-lasting variations. We fitted three models to the generated PSDs, namely a pure power law (PL), a PL with Poisson noise (PLC), and a smoothly broken PL (SBPL). The better model was chosen based on the Akaike Information Criterion \citep[$AIC_c$; ][]{Akai74} which we evaluated via the difference, $\Delta_i=AIC_{c,i}-AIC_{c,\rm min}$, between the $AIC_c$ of the $i$-th model and the minimal value ($AIC_{c,\rm min}$). If $\Delta_i<2$, then both models are equally good and we chose the pure PL model as the adequate one since it is simpler. \\

\textbf{Wavelet scalogram} \citep{Lenoir2018} is a two-dimensional time-frequency representation of the energy-density map showing the temporal localisation of a frequency present in the signal and allowing us to study the local components and their time evolution. The significance testing to search for QPOs at the level $\geq3\sigma$ was employed within this method.\\

\textbf{ARMA and CARMA modelling}: the autoregressive moving average process \citep[ARMA;][]{Scargle1981} and the continuous-time ARMA process \citep[CARMA;][]{kelly2014} are stochastic processes applied to detect different types of variability in the data, to uncover QPOs, and to determine the variability-based classification of the astrophysical sources. In the case of the CARMA processes, a PSD can be composed of a number of zero-centered Lorentzians, which define breaks, while the non-zero-centered Lorentzians are used to model QPOs. Moreover, the CARMA modelling allows us to handle irregular sampling and error measurements.\\

\textbf{Hurst exponent} \citep{Hurs51} measures the statistical self-similarity of a time series. The self similarity is connected to a long range dependence, referred to as \textit{memory}, of a process via the autocorrelation function. The properties of $H$ can be summarized as follows: $H$ takes values between 0 and 1; $H=0.5$ is for an uncorrelated process (white noise or Brownian motion); if $H>0.5$, than a process is persistent, i.e. exhibits long-term memory; while in the opposite case, if $H<0.5$, one deals with an anti-persistent (short-term memory) process.\\

\textbf{The $\mathcal{A-T}$ plane} \citep{tarnopolski2016} is used to differentiate various types of coloured noise. The plane consists of the fraction of \textit{turning points} ($\mathcal{T}$) to verify the noisiness of a time series and the \textit{Abbe value} ($\mathcal{A}$) which quantifies the smoothness of a time series. If $\mathcal{T}$ is asymptotically equal to 2/3, the time series constitutes a purely random time series or white noise. A process with $\mathcal{T}>2/3$ is more noisy than white noise. For $\mathcal{T}<2/3$, a process is less noisy than white noise.

\section{Results}

We analysed the \textit{Fermi}-LAT $\gamma$-ray LCs of 11 blazars, six FSRQs and five BL Lacs, employing a number of techniques. We found the following results.

\begin{figure}[!h]
\centering
\includegraphics[width=0.9\textwidth]{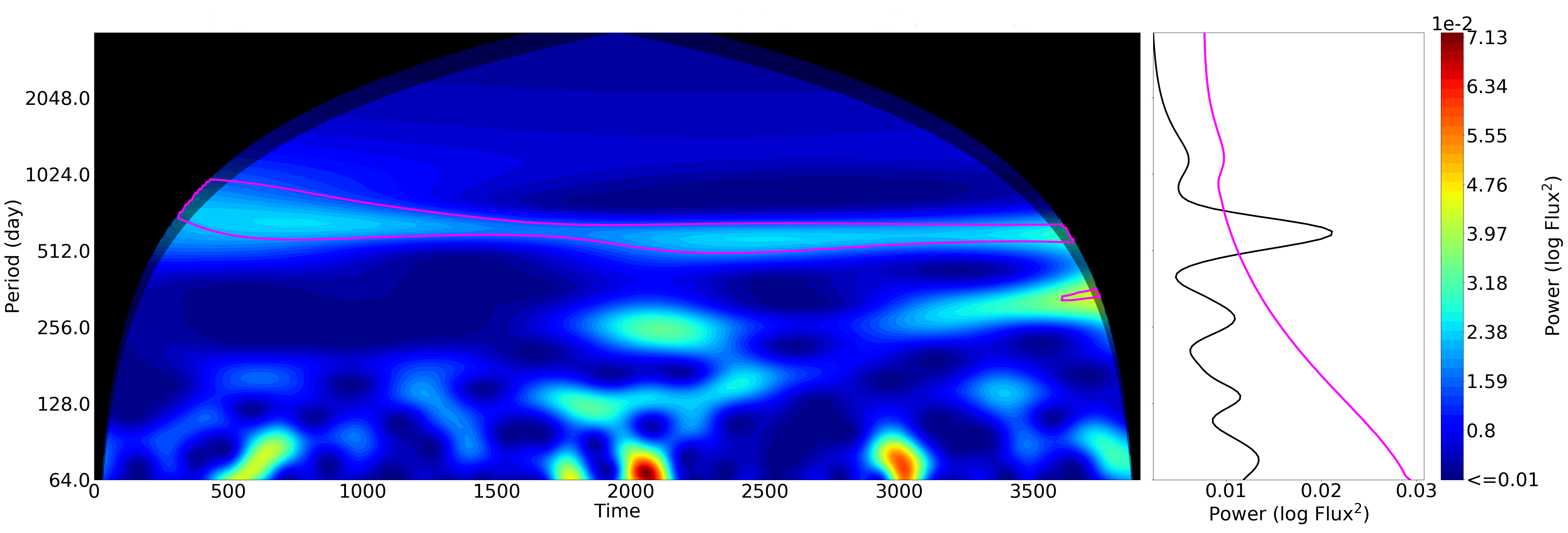}
\caption{Wavelet scalogram of the 7-day binned LC of PKS~2155-304. Right panel displays the global wavelet periodogram. The magenta contours in the scalogram and the magenta line in the periodogram denote 3$\sigma$ local and global confidence levels, respectively. The shadowed area shows the cone of influence, i.e. the erroneous zone.}
\label{Scalogram}
\end{figure}

\begin{figure}[!h]
\centering
\includegraphics[width=0.7\textwidth]{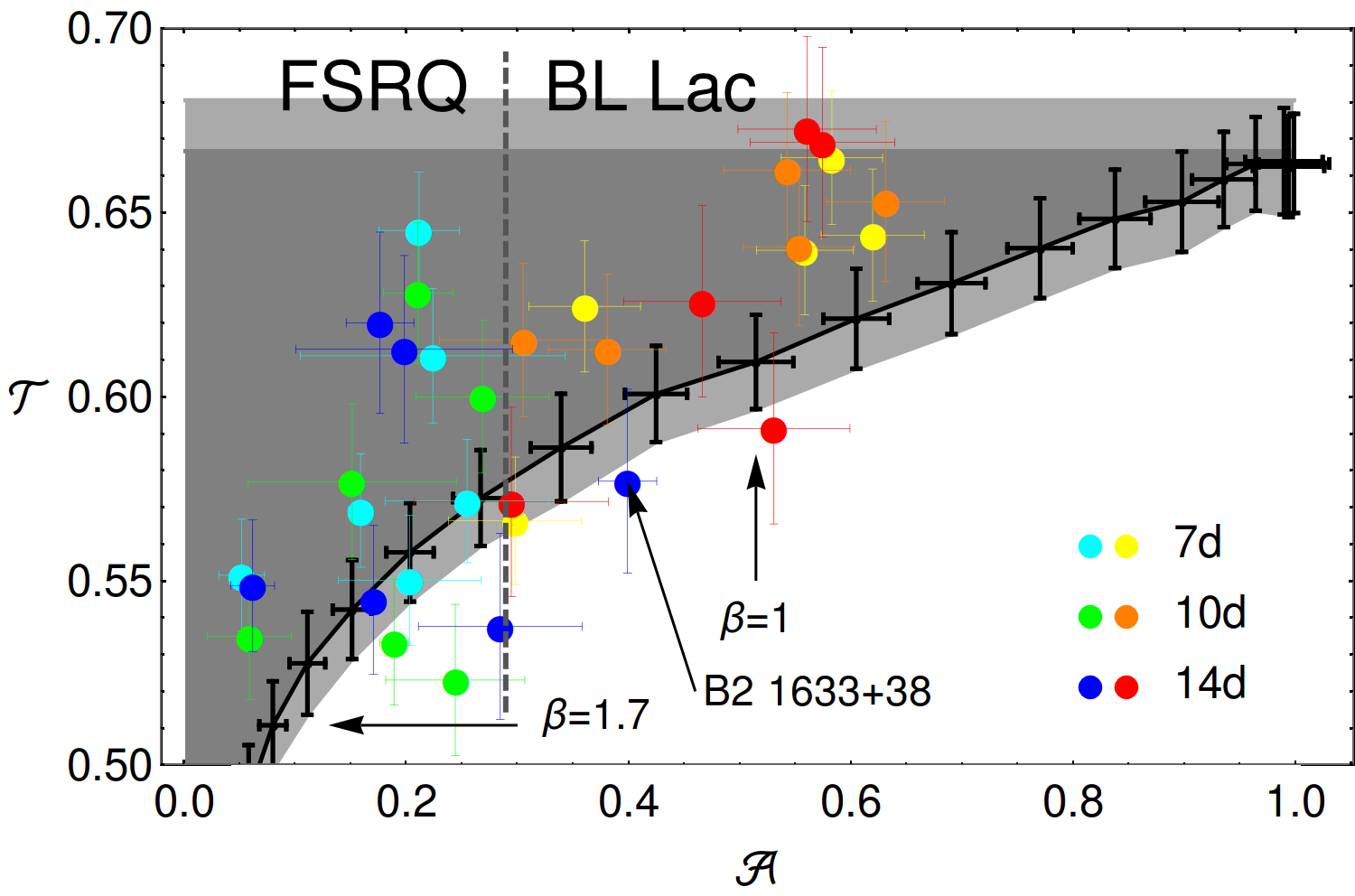}
\caption{Locations in the $\mathcal{A}-\mathcal{T}$ plane of blazars from our sample. The FSRQs are denoted with blue and green colors, while BL Lacs are marked with with yellow, orange, and red. The separation between FSRQs and BL Lacs is denoted with the dashed vertical gray line. The dark gray area is the region between the pure PL line and $\mathcal{T}=2/3$, while the light gray regions represent the error bars of the simulations. }
\label{fig_AT_separation}
\end{figure}

\begin{enumerate}

\item The $\beta$ values calculated with the Fourier spectra and the LSP are consistent with each other for the majority of blazars; however, we noticed a discrepancy for 3C~454.3. In this case, the Fourier PSD is described by a pure PL, while the LSP is fitted better by a PLC. The former PSD is flatter than the latter. The fit of the SBPL model was not competing in any of the cases under consideration. Overall, the shapes of PSDs indicate a coloured noise with $1\lesssim\beta\lesssim 2$, i.e. between pink and red noise. We suggest that each object can be treated as realisation of one stochastic process underlying the observed variability.

\item The only significant ($\geq3\sigma$) QPO we found using the wavelet scalograms is the well-known QPO in PKS~2155$-$304, with a period of $612\pm 42$~days (Figure~\ref{Scalogram}). Moreover, we noticed a QPO candidate in B2~1633+38 data, evolving from $P\sim 500$~days to $P>1000$~days, and in PKS~0716+714 at $P\gtrsim 1000$~days, lasting 2 to 3 cycles only. These objects require additional observations to actually conclude whether a QPO exists in their data. We do not find significant QPOs in the studied LCs of the remaining blazars in our sample. 


\item ARMA and CARMA models suggest breaks in the PSDs at time scales of a few hundred days in all blazars in the sample but 3C~454.3 and B2~1520+31. We searched for the best CARMA$(p,q)$ model with $1\leqslant p\leqslant 7$ and $0\leqslant q\leqslant 6$, $q < p$. We obtained the orders $(1,0)$ or $(2,1)$ for the majority of cases. In general, FSRQs are described with the former model, while the latter represents BL Lacs.  We do not observe any additional features in the data of the analysed objects. 

\item The Hurst exponents are $>0.5$ for the majority of BL Lacs in our sample, indicating the presence of long-term memory. The FSRQs swing back and forth between $H\lesssim 1$ and $H\gtrsim 0$. Only 3C~454.3 keeps $H<0.5$, being an anti-persistent system. Also Mrk~421 and PKS~0716+714 are oscillating in the entire range of the $H$ limit. This evolution behaviour does not allow us to formulate an unambiguous claim about the persistence of these objects.

\item The FSRQs are characterised by lower values of $\mathcal{A}$ than BL Lacs and these two classes of blazars are clearly separated on the $\mathcal{A-T}$ plane (Figure~\ref{fig_AT_separation}). This was earlier discovered by \cite{zywucka2020} in the I-band optical LCs of blazars and blazar candidates behind the Magellanic Clouds \citep{zywucka2018}. This finding shows that the flux changes are different for the two blazar classes, thus they should be driven by different physical mechanisms or take place in different blazar components. The separation allows us to distinguish blazar classes based on LCs without including multiwavelength, polarimetric, and spectroscopic properties. Furthermore, the location in the $\mathcal{A}-\mathcal{T}$ plane indicates properties in the structure of LCs, which are not revealed by other methods used in this work. 

\end{enumerate}

\section{Conclusions}

In our research, we considered a stochastic description to model the variability of blazars. All blazars in our sample are characterized by long timescales consistent with a conclusion that their variability originates in the accretion disk. The timescales also point out the physical processes responsible for $\gamma$-ray production, i.e. external Compton in the case of FSRQs and synchrotron self-Compton for the BL Lacs. The detailed elaboration on results and conclusions is presented in \cite{Tarnopolski2020}.

\acknowledgments
The work of N.\.{Z}. is supported by the South African Research Chairs Initiative (grant no. 64789) of the Department of Science and Innovation and the National Research Foundation\footnote{Any opinion, finding and conclusion or recommendation expressed in this material is that of the authors and the NRF does not accept any liability in this regard.} of South Africa. M.T. acknowledges support by the Polish National Science Center (NSC) through the OPUS grant No. 2017/25/B/ST9/01208. V.M. is supported by the NSC grant No. 2016/22/E/ST9/00061. J.P.-G. acknowledges financial support from the State Agency for Research of the Spanish MCIU through the ''Center of Excellence Severo Ochoa'' award to the Instituto de Astrof\'isica de Andaluc\'ia (SEV-2017-0709) and from Spanish public funds for research under project ESP2017-87676-C5-5-R.


\end{document}